# Analysis of defect modes in one-dimensional photonic crystals with two defects

A.O. Kamenev[1,2*], N.A. Vanyushkin[1], I.M. Efimov[1] and A.H. Gevorgyan[1]

[1]Far Eastern Federal University, Institute of High Technologies and Advanced Materials, 10 Ajax Bay, Russky Island, Vladivostok 690922, Russia
[2]Institute of Automation and Control Processes, Far East Branch, Russian Academy of Sciences, 5, Radio str., Vladivostok 690041, Russia
*E-mail: kamenev.ao@dvfu.ru

**Abstract**

In this paper, for the first time, analytical expressions for determination of the wavelength of defect modes (DMs) in one-dimensional (1D) photonic crystals (PCs) with two defect layers (DLs) were obtained from the condition of zero reflection. Analytical formulas on the conditions for the merging of two DMs and on the DMs with zero value of the reflection coefficient were obtained. Different typical 1D PC structures with two DLs were considered and compared, including their DMs behavior. The distribution of electromagnetic field within the first photonic bandgap (PBG) was analyzed for several cases of DM merging and for the different orders of DMs. The results of this research permit the simplification of the analysis of optical sensors and filters based on 1D PCs with two DLs, as well as a more comprehensive understanding of the DM behavior within the PBGs of such PC structures. Furthermore, this paper helps to explain the results obtained by other researchers earlier in this field, too.

**Keywords:** photonic crystals with two defect layers, photonic band gap, wave phase, field localization, defect modes, modes merging

**1. Introduction**

With the development of optics and photonics, research into the possibilities of photonic crystals (PCs) is becoming increasingly popular. These structures are artificially created or self-organizing nano-structures with periodic refractive index changes in one (1D PC), two (2D PC) or three (3D PC) spatial dimensions [1-2]. A well-known property of such structures is their ability to control the light passing through them, among other things, due to the presence of so-called photonic bandgap (PBG), which are wavelength ranges where light propagation through PCs is impossible [3-6]. The presence of PBGs provides many applications for 1D PCs, including polarizers [7], multichannel optical switches [8], filters [9], absorbers [10], etc. When the periodicity of 1D PCs is disrupted by the addition of a defect layer (DL), a narrow defect mode (DM) appears within their PBG. In 1D PC, the DM represents a resonant peak in the transmission spectra. The number of DMs within the PBG is determined by the number of DLs in the structure of the PC and their optical thicknesses. The parameters of the DLs directly influence the position of the DM in the PC spectrum, which is the basis for the operation of 1D PC-based sensors. Over the past few decades, many different optical sensors based on 1D PCs with DLs have been developed for a wide range of applications. These include gas sensors [11-13], sensors for liquids [14], temperature sensors [15], biochemical sensors [16] and sensors for the detection of harmful viruses and bacteria in the human body [17]. Furthermore, low-threshold lasers on DMs can be created on the basis of various PCs [18-22]. Additionally, many papers have been published on the optimization of 1D PCs and their properties [23-28]. At the same time, most researchers deal with 1D PCs with a single DL in the center of the structure due to the simplicity of their analysis. For this reason, the number of works concerning PCs with two or more DLs is rather small. The presence of several DMs in the spectrum allows extending the application possibilities of such PCs. One of the obvious applications of such structures is their promising potential as optical filters, due to the possibility of manipulation of two or more DMs independently by varying the number and thickness of DLs in the PC structure [29, 30] or by tuning their properties by external (electrical, magnetic, light, mechanical, temperature, etc) field. Other researchers have noted the following perspectives of application of one-dimensional PCs with two DLs in the structure. In [31], the shape change of Gaussian laser pulses transmitted through 1D PC with two DLs was investigated in both linear and nonlinear cases. In [32], the impact of cubic optical nonlinearity on the behavior of normally incident light traversing a magnetic PC with two DLs was investigated. In [33-35], the influences of the arrangement of DLs with respect to each other on the position of DMs in the transmission spectra were investigated. In [36, 37], the advantages of 1D PCs with two DLs for fabrication of ultrafast all-optical switching devices and for broadband energy localization under the action of elastic waves were shown. In [38], the reflection spectra of a cholesteric liquid crystal layer with two DLs were studied, and it was shown that self-organizing PCs could be used for adaptive optical devices consisting of a cholesteric liquid crystal layer with two DLs within it. Finally, researches [39, 40] demonstrated the promising potential of one-dimensional dual-defect-mode PCs for the development of temperature sensors and sensors for biomedical applications. These works also highlight the importance of the investigation of structures with two DLs.

In general, numerical calculations are used to find the DM wavelength and other parameters of sensors based on defective PCs. However, there are several publications in which analytical expressions for the DM wavelength have been obtained [41-45], but only for the structure with one defect. Some analytical expressions for resonances in the structure with two DLs were obtained from the generalized temporal coupled mode theory [46]. However, this approach in the case of PCs with two DLs is only phenomenological, and the authors do not provide a comparison with numerical results, and do not show a way to calculate phenomenological variables in the case of such PCs. Usually, the most common method is to consider either the condition of localization of the DM within the resonator (field attenuation at a distance from the DL) or the condition of zero reflection. Nevertheless, to the best of our knowledge, analytical expressions for determination the position of the DMs within the PBGs in 1D PCs with two DLs have not been previously obtained. Some cases of DM merging have been considered in certain works [33, 34, 47], but there has been no comprehensive investigation of this topic and only the case of mode merging at a perfectly reflecting central mirror of the PC was considered. Consequently, the present paper is devoted to solving this problem. The structure of this article is as follows: section 1 is the introduction part. Section 2 presents the theoretical basis of the well-known transfer matrix method, which we used for numerical simulations of various structures and analytical investigation of 1D PCs with two DLs. Section 3 presents the derivation of the DM wavelengths. Also, in this section the condition for the merging of two DMs on the reflection spectrum is obtained. All analytical results are illustrated by the corresponding numerical results shown in the figures. In the same section, the intensity distribution of the electromagnetic field inside structures with the merging DMs is shown. Section 4 demonstrates the derivation of analytical equations for DMs with zero reflection. Section 5 presents a comparative analysis of the several PC structures with two DLs and illustrates the characteristics of the optical radiation and the field distribution within each structure. Section 6 presents the concluding remarks of the work.

## 2. Theory

Let us consider the reflection and transmission of a plane electromagnetic wave through a 1D binary PC with a double defect (Figure 1) which consists of two DLs with refractive indices $n_{d1}$ and $n_{d2}$ and thicknesses $d_{d1}$ and $d_{d2}$ sandwiched sequentially between three PCs, which are made of $N_1$, $N_2$ and $N_3$ unit cells each and playing the role of mirrors. The unit cell consists of the layer A with refractive index $n_1$ and thickness $d_1$ and the layer B with refractive index $n_2$ and thickness $d_2$. For clarity we will assume that $n_2 > n_1$ throughout the paper. The exact layer order in the unit cell can be ether AB or BA and can be the same or reversed for the three mirrors. The structure without mirror symmetry (MS) containing such layers A and B and two DLs D1 and D2, we will denote by the following expression: AB|D1|AB|D2|AB. The regions outside the structure are filled with vacuum ($n_0 = 1$).

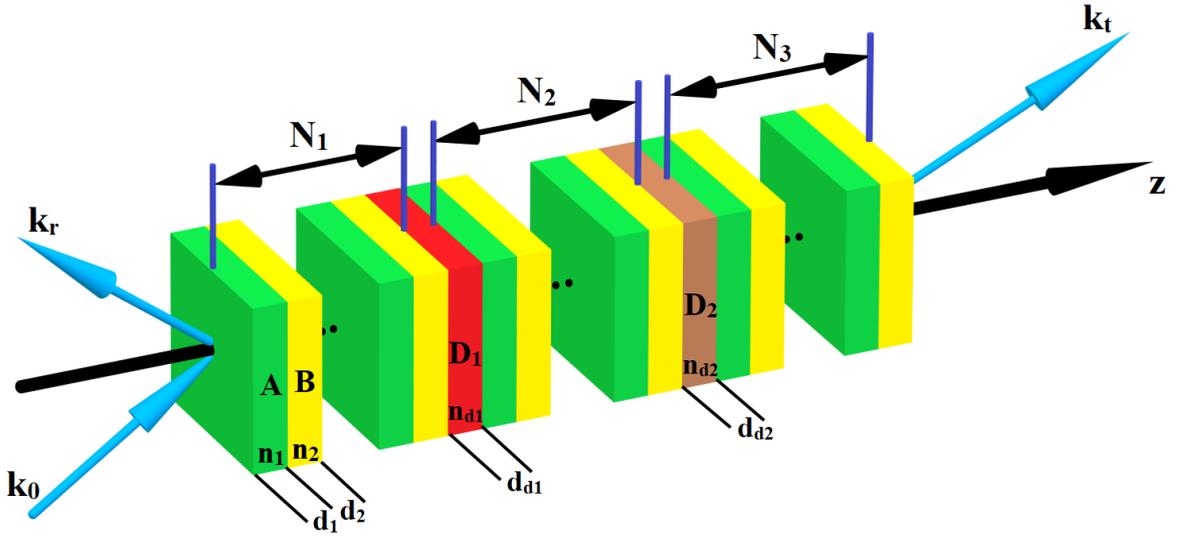

**Figure 1.** Geometry of the problem using the example of a structure without MS (AB|D1|AB|D2|AB). Here $k_0$, $k_r$, $k_t$ are the wave vectors of the incident, reflected and transmitted waves, respectively.

For numerical calculations of the reflection and transmission spectra of light through the PC, we used the well-known transfer-matrix method [48]. The transfer matrix relates the field amplitudes at the different interfaces and can be represented for the *j*-th layer in the following form:

$$M_j = \begin{pmatrix} \cos k_j d_j & \frac{-i}{p_j} \sin k_j d_j \\ -i p_j \sin k_j d_j & \cos k_j d_j \end{pmatrix}, \quad (1)$$

where $k_j = \frac{\omega}{c} n_j \cos\theta_j$, $d_j$ is the thickness of the *j*-th layer, $\theta_j$ is the angle of refraction which is determined from Snell's law as: $n_j \sin\theta_j = n_0 \sin\theta_0$, $n_0$ is the refractive index of the external medium from where the wave is incident, $\theta_0$ is angle of incidence. Also $p_j = n_j \cos\theta_j$ for *s*-wave and $p_j = (1/n_j)\cos\theta_j$ for *p*-wave. Then the transfer matrix *m* of a PC consisting of N layers can be obtained by multiplying the transfer matrices of all layers:

$$m = M_1 M_2 \ldots M_{N-1} M_N = \begin{pmatrix} m_{11} & m_{12} \\ m_{21} & m_{22} \end{pmatrix}, \quad (2)$$

The complex transmission and reflection coefficients for *s*- and *p*-waves are given as:

$$t_N = \frac{2p_0}{(m_{11}+ m_{12}p_0)p_0+ (m_{21}+m_{22}p_0)}, \quad (3)$$

$$r_N = \frac{(m_{11}+ m_{12}p_0)p_0- (m_{21}+m_{22}p_0)}{(m_{11}+ m_{12}p_0)p_0+ (m_{21}+m_{22}p_0)}, \quad (4)$$

Here $p_0 = n_0\cos\theta_0$ for *s*-wave and $p_0 = (1/n_0)\cos\theta_0$ for *p*-wave.

On the other hand, the transfer matrix of any structure relating the amplitudes of waves travelling in different directions without absorption or gain can be expressed through the reflection $r$ and transmission $t$ coefficients of that structure as follows:

$$M = \begin{pmatrix} \frac{1}{t} & \frac{r^*}{t^*} \\ \frac{r}{t} & \frac{1}{t^*} \end{pmatrix}. \quad (5)$$

The DL transfer matrix without taking into account the reflection from the boundaries will then take the following form:

$$M_d = \begin{pmatrix} e^{-i\varphi} & 0 \\ 0 & e^{i\varphi} \end{pmatrix}, \quad (6)$$

where
$$\varphi = \frac{2\pi}{\lambda}n_d d_d. \quad (7)$$

$\varphi$ is the change of phase of the wave at a single passage through the DL. It should be noted that the phase change will vary for each DL. Consequently, for two DLs in the PC structure, we can express this as follows: $\varphi_1 = \frac{2\pi}{\lambda}n_{d1}d_{d1}$, $\varphi_2 = \frac{2\pi}{\lambda}n_{d2}d_{d2}$.

## 3. Analysis of DM behavior in 1D PC with two DLs

The transfer matrix of a PC with two defects can be generally represented as:

$$m = (M_{01}M_1M_{1d})M_{dd1}(M_{d2}M_2M_{2d})M_{dd2}(M_{d3}M_3M_{30}) = M_I M_{dd1} M_{II} M_{dd2} M_{III}, \quad (8)$$

where $M_1, M_{dd1}, M_2, M_{dd2}, M_3$ are "inner" transfer matrices of PC$_1$, DL$_1$, PC$_2$, DL$_2$ and PC$_3$, i.e. these transfer matrices relate the fields at inner sides of their outer boundaries. $M_{01}$ and $M_{30}$ are "boundary" matrices for the interfaces between the perfect PCs and the external environment, i.e. these transfer matrices relate the fields at different sides of the same boundary. Similarly, the matrices for the boundaries between the perfect PCs and DLs are defined as $M_{1d}, M_{d2}, M_{2d}$ and $M_{d3}$. For convenience, these transfer matrices can be combined into left $M_I$, central $M_{II}$ and right $M_{III}$ mirror' matrices.

If we write equation (8) in explicit form through reflection and transmission coefficients, we obtain:

$$\begin{pmatrix} \frac{1}{T} & \frac{R^*}{T^*} \\ \frac{R}{T} & \frac{1}{T^*} \end{pmatrix} = \begin{pmatrix} \frac{1}{t_I} & \frac{r_I^*}{t_I^*} \\ \frac{r_I}{t_I} & \frac{1}{t_I^*} \end{pmatrix} \begin{pmatrix} e^{-i\varphi_1} & 0 \\ 0 & e^{i\varphi_1} \end{pmatrix} \begin{pmatrix} \frac{1}{t_{II}} & \frac{r_{II}^*}{t_{II}^*} \\ \frac{r_{II}}{t_{II}} & \frac{1}{t_{II}^*} \end{pmatrix} \begin{pmatrix} e^{-i\varphi_2} & 0 \\ 0 & e^{i\varphi_2} \end{pmatrix} \begin{pmatrix} \frac{1}{t_{III}} & \frac{r_{III}^*}{t_{III}^*} \\ \frac{r_{III}}{t_{III}} & \frac{1}{t_{III}^*} \end{pmatrix}. \quad (9)$$

Here $R$ and $T$ are the reflection and transmission coefficients of the whole structure:

$$R = \frac{e^{2i(\varphi_1 + \varphi_2)}r_{\mathrm{III}}t_{\mathrm{I}}t_{\mathrm{II}} + e^{2i\varphi_2}r_{\mathrm{I}}r_{\mathrm{III}}t_{\mathrm{II}}r_{\mathrm{II}}^* t_{\mathrm{I}}^* + (e^{2i\varphi_1}r_{\mathrm{II}}t_{\mathrm{I}} + r_{\mathrm{I}}t_{\mathrm{I}}^*)t_{\mathrm{II}}^*}{t_{\mathrm{I}}^*(e^{2i\varphi_2}r_{\mathrm{III}}t_{\mathrm{II}}r_{\mathrm{II}}^* + t_{\mathrm{II}}^*) + e^{2i\varphi_1}t_{\mathrm{I}}r_{\mathrm{I}}^*(e^{2i\varphi_2}r_{\mathrm{III}}t_{\mathrm{II}} + r_{\mathrm{II}}t_{\mathrm{II}}^*)}. \quad (10)$$

$$T = \frac{e^{i(\varphi_1 + \varphi_2)}t_{\mathrm{I}}t_{\mathrm{II}}t_{\mathrm{III}}t_{\mathrm{I}}^* t_{\mathrm{II}}^*}{t_{\mathrm{I}}^*(e^{2i\varphi_2}r_{\mathrm{III}}t_{\mathrm{II}}r_{\mathrm{II}}^* + t_{\mathrm{II}}^*) + e^{2i\varphi_1}t_{\mathrm{I}}r_{\mathrm{I}}^*(e^{2i\varphi_2}r_{\mathrm{III}}t_{\mathrm{II}} + r_{\mathrm{II}}t_{\mathrm{II}}^*)}. \quad (11)$$

Since the reflection coefficient $R$ is zero for DMs [49], then by demanding $R = 0$ (at the absence of absorption it is also a condition of coherent total transmittance) in Eq. (10) we obtain the characteristic equation containing both phases for both DLs in the PC structure:

$$e^{2i(\varphi_1 + \varphi_2)}r_{\mathrm{III}}t_{\mathrm{I}}t_{\mathrm{II}} + e^{2i\varphi_2}r_{\mathrm{I}}r_{\mathrm{III}}t_{\mathrm{II}}r_{\mathrm{II}}^* t_{\mathrm{I}}^* + (e^{2i\varphi_1}r_{\mathrm{II}}t_{\mathrm{I}} + r_{\mathrm{I}}t_{\mathrm{I}}^*)t_{\mathrm{II}}^* = 0. \quad (12)$$

To further simplify the Eq. (12), we introduce the reflection coefficient of the left mirror when the wave is incident from the right, i.e., from the DL side, as was done in [50, 51]: $\tilde{r}_{\mathrm{I}}^* = -r_{\mathrm{I}} t_{\mathrm{I}}^*/t_{\mathrm{I}}$. For the center mirror, this conversion has a similar form: $\tilde{r}_{\mathrm{II}}^* = -r_{\mathrm{II}} t_{\mathrm{II}}^*/t_{\mathrm{II}}$. By introducing these transformations and multiplying both parts of the Eq. (12) by $r_{\mathrm{II}}$, expression (12) can be transformed into the following form:

$$e^{2i(\varphi_1 + \varphi_2)}r_{\mathrm{III}}r_{\mathrm{II}} - e^{2i\varphi_2}\tilde{r}_{\mathrm{I}}^* r_{\mathrm{III}}|r_{\mathrm{II}}|^2 - e^{2i\varphi_1}\tilde{r}_{\mathrm{II}}^* r_{\mathrm{II}} + \tilde{r}_{\mathrm{I}}^* \tilde{r}_{\mathrm{II}}^* = 0. \quad (13)$$

Finally, we obtain:

$$\left(e^{2i\varphi_1} - \frac{\tilde{r}_{\mathrm{I}}^*}{r_{\mathrm{II}}}|r_{\mathrm{II}}|^2\right)\left(e^{2i\varphi_2} - \frac{\tilde{r}_{\mathrm{II}}^*}{r_{\mathrm{III}}}\right) + \frac{\tilde{r}_{\mathrm{I}}^* \tilde{r}_{\mathrm{II}}^*}{r_{\mathrm{II}} r_{\mathrm{III}}}(1 - |r_{\mathrm{II}}|^2) = 0. \quad (14)$$

The resulting Eq. (14) is very revealing. The first component consists of two multipliers, the first of which is determined only by the parameters of the left cavity ($\tilde{r}_{\mathrm{I}}^*$, $r_{\mathrm{II}}$ and $\varphi_1$), and the second is only by the parameters of the right cavity ($\tilde{r}_{\mathrm{II}}^*$, $r_{\mathrm{III}}$ and $\varphi_2$). The second component of Eq. (14) represents the coupling coefficient between the first and second cavities in the structure and we will denote it as $J = \frac{\tilde{r}_{\mathrm{I}}^* \tilde{r}_{\mathrm{II}}^*}{r_{\mathrm{II}} r_{\mathrm{III}}}(1 - |r_{\mathrm{II}}|^2)$.

From the Eq. (14) on defect phases, we can express the solution for $\varphi_1$ through $\varphi_2$:

$$\varphi_1 = -\frac{1}{2}i \cdot \ln \frac{\tilde{r}_{\mathrm{I}}^*(e^{2i\varphi_2}r_{\mathrm{III}}|r_{\mathrm{II}}|^2 - \tilde{r}_{\mathrm{II}}^*)}{r_{\mathrm{II}}(e^{2i\varphi_2}r_{\mathrm{III}} - \tilde{r}_{\mathrm{II}}^*)} + \pi q, \quad (15)$$

or vice versa:

$$\varphi_2 = -\frac{1}{2}i \cdot \ln \frac{\tilde{r}_{\mathrm{II}}^*(e^{2i\varphi_1}r_{\mathrm{II}} - \tilde{r}_{\mathrm{I}}^*)}{r_{\mathrm{III}}(e^{2i\varphi_1}r_{\mathrm{II}} - |r_{\mathrm{II}}|^2 \tilde{r}_{\mathrm{I}}^*)} + \pi q, \quad (16)$$

where $q \in \mathbb{Z}$.

If the reflection coefficient of the central mirror in Eq. (14) is equal to unity ($|r_{\mathrm{II}}|^2 = 1$), then the coupling coefficient $J$ is zero. In this case, the left and right cavities behave as two independent defective PCs, and there are independent equations for each phase $\varphi_1$ and $\varphi_2$. It can be shown that the solutions of these equations $e^{2i\varphi_1} - \frac{\tilde{r}_{\mathrm{I}}^*}{r_{\mathrm{II}}} = 0$ and $e^{2i\varphi_2} - \frac{\tilde{r}_{\mathrm{II}}^*}{r_{\mathrm{III}}} = 0$ are independent of each other:

$$\varphi_{10} = -\frac{1}{2}i \cdot \ln \frac{\tilde{r}_{\mathrm{I}}^*}{r_{\mathrm{II}}} + \pi q, \quad (17)$$

$$\varphi_{20} = -\frac{1}{2} i \cdot \ln \frac{\tilde{r}_{II}^*}{r_{III}} + \pi q, \tag{18}$$

Each of these solutions is identical to the solution for PC with a single DL obtained in [51]. Consequently, at $|r_{II}|^2 = 1$, the problem of finding solutions for phases of PCs with two DLs can be reduced to the problem of two independent defective PCs. In this case, the independent behavior of the DMs will be observed. When the parameters of one of the DLs are altered, the position of only one of the DMs will change, while the second DM will remain stationary.

The analysis of the Eq. (14) allows us to make the following assumption. If the DMs behave independently (under the condition $J = 0$), then in a structure with two identical defective PCs combined into one PC with two DLs, the wavelengths of both DMs corresponding to each of the DLs should coincide, i.e., the DMs should merge in the spectrum. It is possible to achieve the condition $J = 0$ when $|r_{II}| = 1$, and this can be obtained, for example, by increasing $N_2$. This phenomenon can be analogized to atomic physics, where the energy spectrum of an electron in an atom is discrete. If a system contains two same atoms that are far apart, their energy levels coincide. If such atoms come close enough, their electrons will be able to travel from one atom to another and the atoms become a single system. The energy levels of such a system will be different from the original ones: their splitting will occur. Just like in the case of PCs with two DLs there is a splitting of the DMs positions when the number of layers $N_2$ in the center of the PC decreases. However, when $N_2 \to \infty$, $|r_{II}|^2 \to 1$ and the DMs corresponding to the two DLs begin to behave independently of each other, like atoms at large distances.

The Figure 2 illustrates this behavior, which shows the dependence of the distance $\Delta\lambda$ between two DM peaks within the first PBG in the spectrum of 1D PC on the absolute value of the coupling coefficient $|J|$. This was achieved by varying the number of layers of the central mirror $N_2$. It can be seen that DMs on the spectrum converge up to merging as the value of $|J|$ decreases (as $N_2$ increases). We used two sets of PC parameters: in the first case (red dots), the DMs merge at the center of the PBG, while in the second case (blue dots) they merge at the sufficient distance to the right of the PBG center (about 100 nm). Thus, the DMs merge in defective PCs, where both DMs behave independently of each other (at $|J| = 0$).

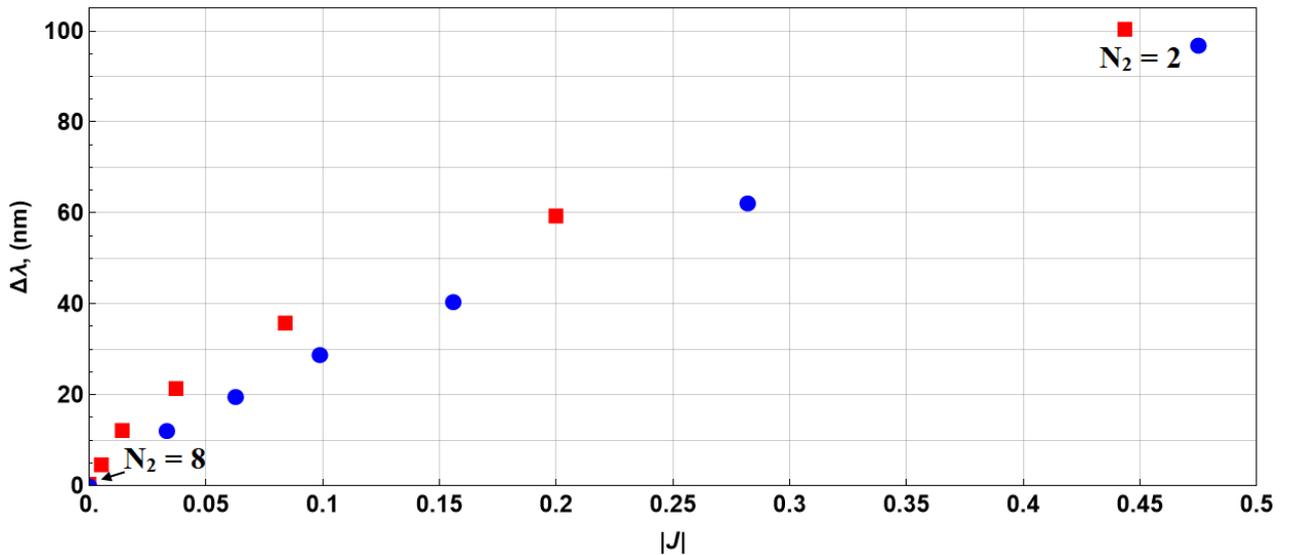

**Figure 2.** Dependence of the distance $\Delta\lambda$ between two DMs within the first PBG in the spectrum of 1D PC with structure AB|D$_1$|BA|D$_1$|AB on the absolute value of the coupling coefficient $|J|$. The structure parameters for the red dots are as follows: $N_1 = N_3 = 4$, $n_1 = 1.5$, $n_2 = 2.4$, $d_1 = 123$ nm, $d_2 = 77$ nm, $d_{d1} = d_{d2} = 0$, $n_{d1} = n_{d2} = 1.33$. For blue dots, the parameters are the same except for the following: $N_1 = N_3 = 3$, $d_{d1} = 63.5$ nm, $d_{d2} = 401.6$ nm, $n_{d1} = n_{d2} = 1.46$. The values of $N_2$ were varied from 8 to 2. Normal light incident on the PC was considered.

However, representing the solutions for the phases in the form of Eqs. (15) and (16) is inconvenient. For example, when the condition $\varphi_1 = \varphi_2$ is required, one must additionally solve Eq. (15) or (16) with respect to $\varphi_1$ or $\varphi_2$ to obtain the final solution. An alternative representation of the solution for the phases $\varphi_1$ and $\varphi_2$ can be obtained by substituting them with the sum of the phases $\varphi_m = \varphi_1 + \varphi_2$ and the phase difference $\Delta\varphi = \varphi_1 - \varphi_2$. Such a substitution will result in the following quadratic equation:

$$e^{2i\varphi_m} - e^{i\varphi_m}\left(\frac{\tilde{r}_{II}^*}{r_{III}}e^{i\Delta\varphi} + \frac{\tilde{r}_I^*}{r_{II}}e^{-i\Delta\varphi}|r_{II}|^2\right) + \frac{\tilde{r}_I^*\tilde{r}_{II}^*}{r_{II}r_{III}} = 0. \tag{19}$$

The solution of the Eq. (19) has the following form:

$$e^{i\varphi_m} = \frac{1}{2}\left(\frac{\tilde{r}_{II}^*}{r_{III}}e^{i\Delta\varphi} + \frac{\tilde{r}_I^*}{r_{II}}e^{-i\Delta\varphi}|r_{II}|^2 \pm \sqrt{\left(\frac{\tilde{r}_{II}^*}{r_{III}}e^{i\Delta\varphi} + \frac{\tilde{r}_I^*}{r_{II}}e^{-i\Delta\varphi}|r_{II}|^2\right)^2 - \frac{4\tilde{r}_I^*\tilde{r}_{II}^*}{r_{II}r_{III}}}\right), \tag{20}$$

which gives

$$\varphi_{m_{1,2}} = -i\cdot\ln\left[\frac{1}{2}\left(\frac{\tilde{r}_{II}^*}{r_{III}}e^{i\Delta\varphi} + \frac{\tilde{r}_I^*}{r_{II}}e^{-i\Delta\varphi}|r_{II}|^2 \pm \sqrt{\left(\frac{\tilde{r}_{II}^*}{r_{III}}e^{i\Delta\varphi} + \frac{\tilde{r}_I^*}{r_{II}}e^{-i\Delta\varphi}|r_{II}|^2\right)^2 - \frac{4\tilde{r}_I^*\tilde{r}_{II}^*}{r_{II}r_{III}}}\right)\right] + 2\pi q_1, \tag{21}$$

where $q_1 \in \mathbb{Z}$.

One can see that we obtained two solutions to the phase, which correspond to the behavior of each DM in the PC spectrum. These solutions for the DL phase are periodical with the period $2\pi$, and the integer number $q_1$ can be associated with the DM order [28].

It is important to pay special attention to the meaning of the sum of the phases $\varphi_m$ from Eq. (21), as well as its difference from sum of the phases, calculated from the DL parameters $\varphi_m' = \frac{2\pi}{\lambda}(n_{d1}d_{d1} + n_{d2}d_{d2})$. On the one hand, the phase $\varphi_m'$ simply shows the change of the phase of the wave when passing the first and second DLs and depends solely on the properties of the DLs and the wavelength of light, but only at some values of $\varphi_m'$ DMs are present at the given wavelength. Usually, we can control $\varphi_m'$ by changing the thickness of the DLs $d_{d1}$ and $d_{d2}$ and choosing a different wavelength $\lambda$. On the other hand, $\varphi_m$ shows what $\varphi_m'$ should be equal to achieve zero reflection coefficients at a given wavelength and mirrors' parameters. Therefore, $\varphi_m'$ represents the actual total phase of the DLs, while $\varphi_m$ is the required total phase at which a DM will be observed. Consequently, wavelengths with $|R|^2 = 0$ correspond to the exact equality $\varphi_m = \varphi_m'$. This means that in the case of complex values of $\varphi_m$ the following equations must be satisfied: $\text{Re}[\varphi_m] = \text{Re}[\varphi_m']$ and $\text{Im}[\varphi_m] = \text{Im}[\varphi_m']$. This leads to the necessity of an imaginary component in the refractive indices $n_{d1}$ and $n_{d2}$. In the case of real $n_{d1}$ and $n_{d2}$ we may encounter $\text{Im}[\varphi_m] \neq \text{Im}[\varphi_m']$ and then the reflection coefficient at the DM wavelength will be different from zero, i.e., the DM amplitude decreases.

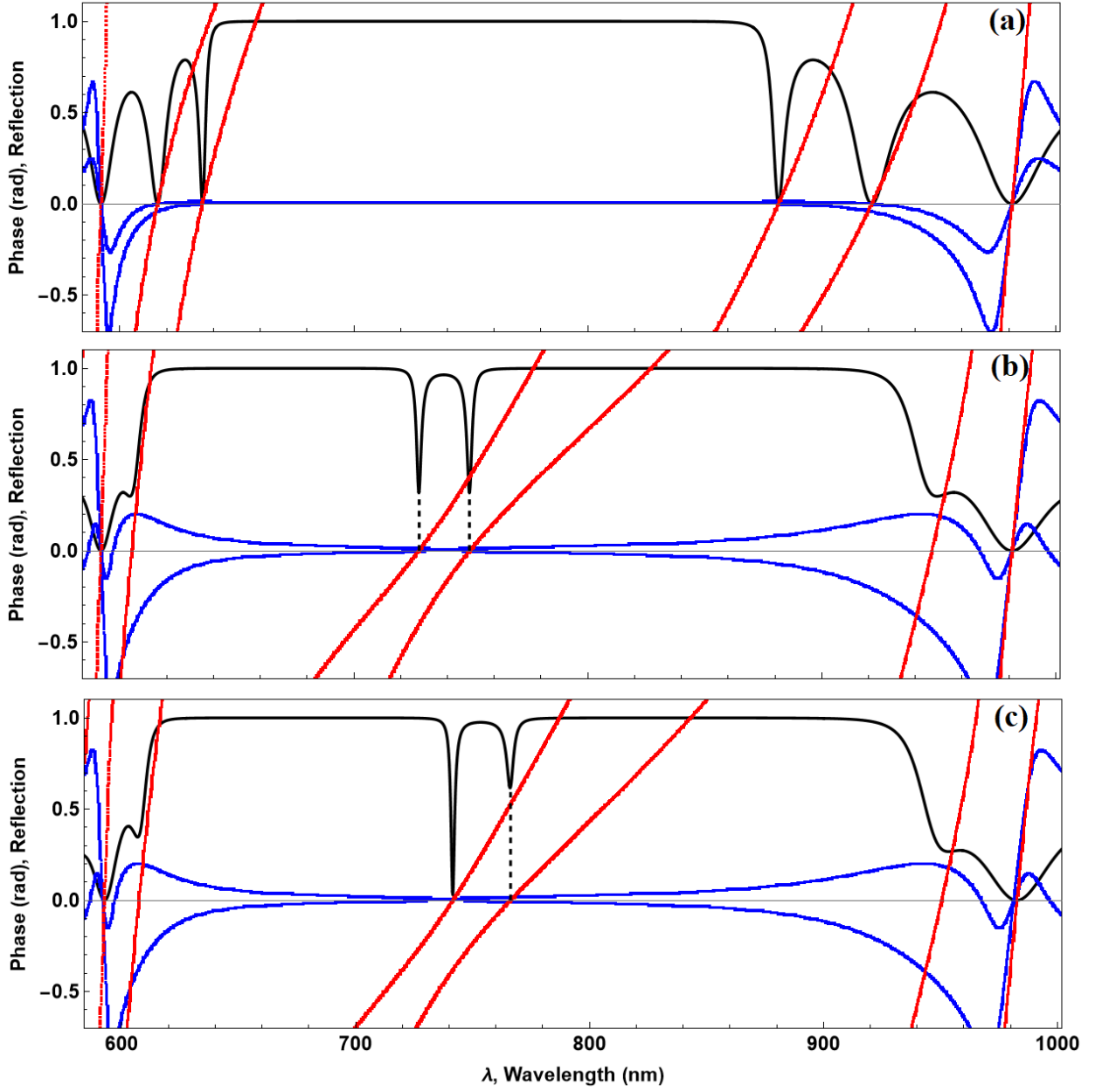

**Figure 3**. Spectra of Re[$\varphi_m - \varphi_m'$] (red), Im[$\varphi_m$] (blue), and reflection spectrum $|R|^2$ (black) for structure (AB|D$_1$|AB|D$_2$|AB) (a) and structure (AB|D$_1$|BA|D$_2$|AB) (b, c). The parameters of the structures are $N_1 = N_2 = N_3 = 5$, $n_1 = 1.5$, $n_2 = 2.4$, $d_1 = 123$ nm, $d_2 = 77$ nm, $d_{d1} = d_{d2} = 0$ (a, b), $d_{d1} = d_{d2} = 11$ nm (c), $n_{d1} = n_{d2} = 1.33$. The normal incidence of light on the PC was considered.

These conclusions are illustrated in Figure 3, which shows the spectra of the real and imaginary parts of the total phase $\varphi_m$ from Eq. (21) for a structure without mirror symmetry (as shown in Figure 1 (a)) and a structure (AB|D$_1$|BA|D$_2$|AB), with the layers of the central mirror arranged in reverse order (b, c), as well as the reflection spectrum $|R|^2$ when $d_{d1} = d_{d2} = 0$ nm (a, b) and $d_{d1} = d_{d2} = 11$ nm (c). Figure 3a clearly shows that all wavelengths with $|R|^2 = 0$ satisfy the equations $\varphi_m = \varphi_m'$ and the Im[$\varphi_m$] = Im[$\varphi_m'$] = 0. From the comparison of Figures 3b and 3c, we can also see that all wavelengths at which $|R|^2 = 0$ necessarily correspond to strict equality of the real and imaginary parts of the total phase: Im[$\varphi_m$] = Im[$\varphi_m'$] = 0 and Re[$\varphi_m$] = Re[$\varphi_m'$] = 0. However, for both DMs $|R|^2 \neq 0$, which is due to the difference between the imaginary total phases at these wavelengths: Im[$\varphi_m$] ≠ Im[$\varphi_m'$]. Figure 3c also shows that the reflection coefficients of the two DMs are different due to the increasing value of $\text{Im}[\varphi_m]$ as we move away from the center of the PBG, while the value of $\text{Im}[\varphi_m']$ is zero everywhere. Thus, when there is difference between the imaginary part of the total phase $\text{Im}[\varphi_m]$, calculated from Eq. (21), and the imaginary part of the total phase $\text{Im}[\varphi_m']$, calculated from the PC parameters, the amplitude of the DM reflection coefficient increases.

We are also interested in the conditions for the merging of two DMs in the spectrum of a 1D PC with two DLs. As previously stated, the majority of researchers consider the influence of the number of layers $N_2$ in the central mirror of such PCs on the position of DMs in the spectrum. Their conclusions demonstrate that with increasing $N_2$, the DMs converge towards each other until they merge. However, this is only a particular case as illustrated in Figure 2. In order to identify a more general condition for the merging of the DMs, we can analyze the solutions to the total phase (21). These solutions differ from each other only by the sign of the square root. This means that if this root is equal to 0, the two solutions to the total phase will be identical, thereby indicating that the modes merge at a specific wavelength. In this case:

$$\left(\frac{\tilde{r}_{II}^*}{r_{III}}e^{i\Delta\varphi} + \frac{\tilde{r}_I^*}{r_{II}}e^{-i\Delta\varphi}|r_{II}|^2\right)^2 - \frac{4\tilde{r}_I^*\tilde{r}_{II}^*}{r_{II}r_{III}} = 0. \tag{22}$$

The solution of this equation with respect to $\Delta\varphi$ after simplifications can be written in the following form:

$$\Delta\varphi = \frac{1}{2}i \cdot \ln\left[\frac{r_{II}\tilde{r}_{II}^*}{r_{III}\tilde{r}_I^*}\frac{2-|r_{II}|^2\pm2\sqrt{1-|r_{II}|^2}}{|r_{II}|^4}\right] + \pi q_2, \tag{23}$$

where $q_2 \in \mathbb{Z}$.

This solution can be further simplified by expanding the reflection coefficients into amplitude and phase:

$$r = |r|e^{i\rho}, \tag{24}$$

Then the final solution to $\Delta\varphi$ will take the following form:

$$\Delta\varphi = \frac{1}{2}(\tilde{\rho}_{II} - \rho_{II} + \rho_{III} - \tilde{\rho}_I) + \frac{1}{2}i \cdot \ln\left[\frac{2-|r_{II}|^2\pm2\sqrt{1-|r_{II}|^2}}{|r_{II}|^2 \cdot |r_{III}| \cdot |r_I|}\right] + \pi q_2. \tag{25}$$

A detailed analysis of the obtained solution for the phase difference (25) reveals a number of significant conclusions. It can be observed that there are infinite number of variants (types of PC structure) in which the DMs on the spectrum will merge. Furthermore, this phenomenon does not occur only at a large number of layers in the central PC. The main condition for this is the satisfaction of Eq. (25) for the phase difference. Obtained solution is complex in the general case. Nevertheless, if the refractive indices of DLs are real, then the phase difference $\Delta\varphi' = \frac{\pi}{\lambda}(n_{d1}d_{d1} - n_{d2}d_{d2})$, calculated from the structure parameters, will never take complex values. Consequently, in this case, the equality $\Delta\varphi = \Delta\varphi'$ will be possible only if the phase difference $\Delta\varphi$ calculated by Eq. (25) is also real. And this requires the following:

$$\ln\left[\frac{2-|r_{II}|^2\pm2\sqrt{1-|r_{II}|^2}}{|r_{II}|^2|r_{III}||r_I|}\right] = 0. \tag{26}$$

Upon analysis of Eq. (26), it can be concluded that in the case of a plus sign in front of $2\sqrt{1-|r_{II}|^2}$, the numerator of the logarithm argument is always greater than unity and the denominator is always smaller than unity, due to which the logarithm argument is always greater than unity and hence Eq. (26) has no solutions. This is why we should leave only a minus sign. In this case, we denote the argument of the logarithm by ξ:

$$\xi = \frac{2-|r_{II}|^2-2\sqrt{1-|r_{II}|^2}}{|r_{II}|^2|r_{III}||r_I|}. \qquad (27)$$

Then, the numerator of the logarithm is always less than one and Eq. (26) has a solution.

It is evident that as $|r_{II}|^2 = 1$, the argument of the logarithm is inversely proportional to the reflection coefficients of the right and left layers of the PC. Consequently, Eq. (27) can be expressed as follows:

$$|r_{III}||r_I| = 1, \qquad (28)$$

which has a solution only at $|r_{III}| = 1$ and $|r_I| = 1$. This is a trivial case when all three mirrors in the PC completely reflect radiation.

The satisfaction of condition (27) leads to DM merging, which is demonstrated in Figure 4. Figure 4a shows the spectra of the real and imaginary parts of total phase $\varphi_m$, the reflection spectrum $|R|^2$, as well as the wavelength dependence of $\xi$ for the structure AB|D₁|AB|D₂|BA, which differs from the structure in Figure 1 by the order of layers in the right mirror. It can be seen that both DMs in the spectrum merge into one single DM at the point where the real phases intersect. In addition, at this wavelength $\xi = 1$, while for all other wavelengths $\xi < 1$. Figure 4b shows a close-up of the intersection of the real and imaginary total phase $\varphi_m$ in the wavelength region of the center wavelength of the first PBG for this structure. This figure shows that the wide bandwidth within the PBG arises due to the presence of four DMs in it simultaneously, two of which merge in the center of the PBG. This conclusion can be made from the behavior of the real part of the total phase $\varphi_m$, but the reflection spectrum is less informative and does not allow us to make this conclusion clearly. Thus, the phase behavior shows the behavior of DMs more clearly than the reflection spectrum.

Here we proceed to analyze the electromagnetic field inside the PC structure with two DLs under the condition of DM merging. Figure 5a illustrates the distribution of the field intensity $|E|^2$ (shown in red) within the 1D PC structure as in Figure 4 (shown in blue) at the DM merging wavelength (approximately 1515 nm). Figure 5b depicts the field intensity $|E|^2$ as a function of wavelength $\lambda$ and coordinate $z$ within the structure. The same method as in our work [25] was used for these calculations. Figure 5a illustrates that in a structure with two DLs, the field intensity $|E|^2$ is located in the regions of two DLs at the DM merging wavelength within the PBG. The observed difference in the field distribution in the areas of the two DLs is possibly due to the difference in their thicknesses and boundary layers. As can be seen from Figure 5b, strong light localization occurs at the merging wavelength of the DM (approximately 1515 nm) as well as at the edges of the PBG, and strong light localization occurs in both DLs of the PC at the wavelengths of the DMs within the PBG. In addition, an interesting conclusion can be drawn about the difference in the coordinates of field localization within and outside the PBG. At DMs within the PBG, the field is localized mostly around the DLs, while outside the PBG it is localized in the center of the structure for the first edge mode.

Figures 6 and 7 also demonstrate the case of merging of two DMs for a different structure at the same wavelength, but the field intensity reaches very large values (up to 15000). This is due to the large optical contrast of the DLs of the PC. The structure parameters are as follows: (AB|D$_1$|AB|D$_2$|BA), $N_1 = N_3 = 7$, $N_2 = 14$, $n_1 = 1.3$, $n_2 = 2.6$, $d_1 = 233.333$ nm, $d_2 = 116.667$ nm, $d_{d1} = 679.104$ nm, $d_{d2} = 452.736$ nm, $n_{d1} = n_{d2} = 1.34$. The field intensity localization regions for the first and second DLs differ both in magnitude and number of peaks. This is also due to the different DLs in the structure. However, for this structure, the amplitude of the DM at the merging point is not maximal. This is due to the fact that the total imaginary phase actually takes values different from zero in the whole region of the first PBG. It is worth noting that in a special case it is possible to obtain a PC structure, where the DMs will merge and the dependencies of real and imaginary parts of both solutions for total phase on wavelength will converge in the whole wavelength region of the first PBG. This makes it possible to manipulate the position of the DMs merging point by changing the PC parameters, for example, the thicknesses of DLs. Such case corresponds to the condition (28).

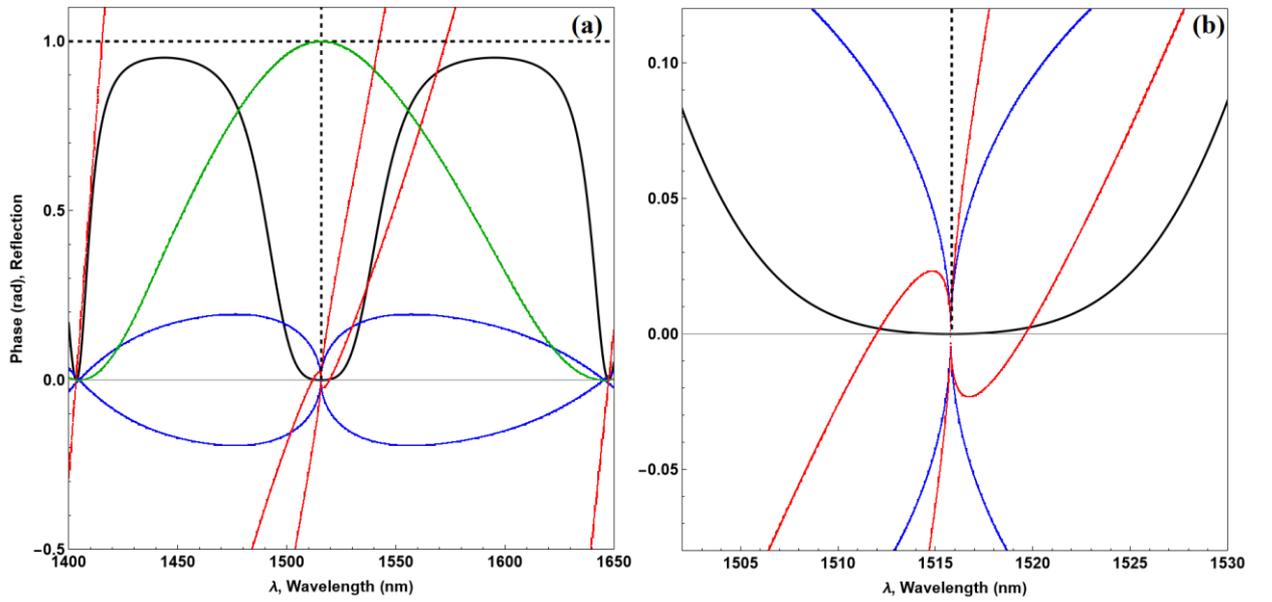

**Figure 4**. Spectra of Re[$\varphi_m - \varphi_m'$] (red), Im[$\varphi_m$] (blue) and reflection spectrum $|R|^2$ (black) for structure (AB|D$_1$|AB|D$_2$|BA) (a); area of the center of the first PBG in the enlarged scale (b). The parameters of the structure are $N_1 = N_3 = 9$, $N_2 = 14$, $n_1 = 1.8$, $n_2 = 2.0$, $d_1 = 210.5$ nm, $d_2 = 189.5$ nm, $d_{d1} = 248.6$ nm, $d_{d2} = 0$, $n_{d1} = n_{d2} = 1.52$. The normal incidence of light on the PC was considered.

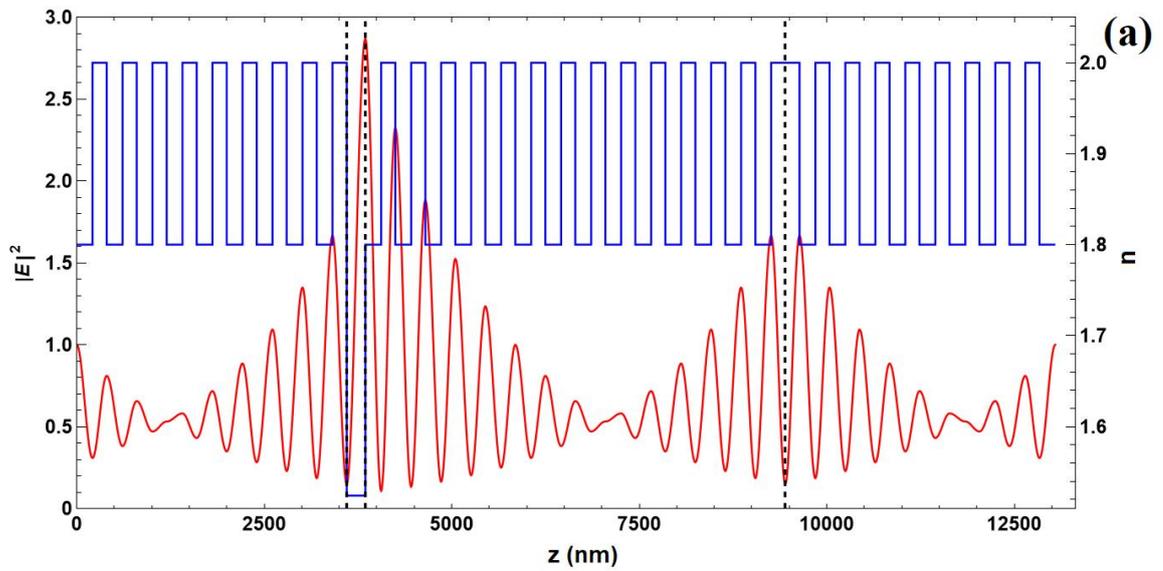

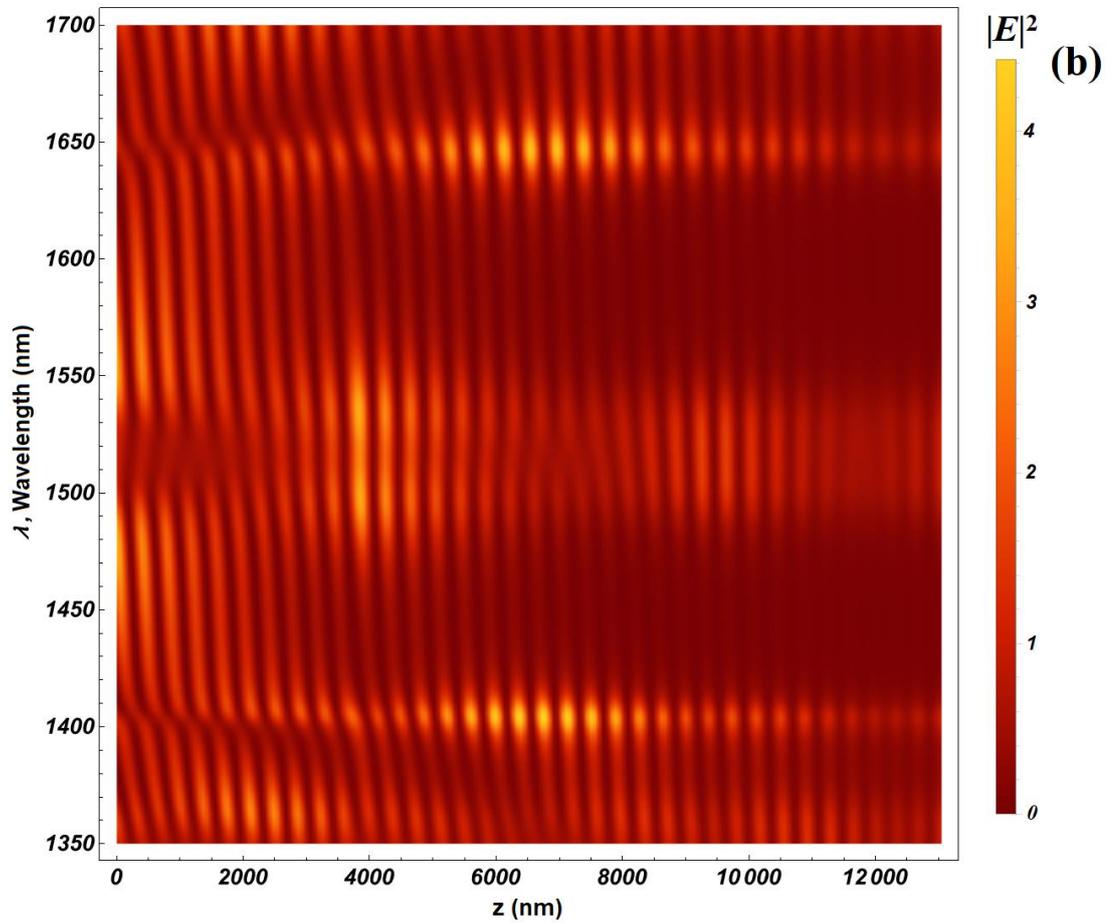

**Figure 5.** Electromagnetic field intensity $|E|^2$ inside the 1D PC with the structure (AB|D$_1$|AB|D$_2$|BA), where $z$ is the coordinate inside the PC (red), and the refractive index (blue) (a); evolution spectrum of the field intensity $|E|^2$ when the wavelength $\lambda$ and the coordinate inside the structure $z$ change (b). The parameters of the structure are the same as in Figure 4. Dashed lines show the boundaries between PC and DL.

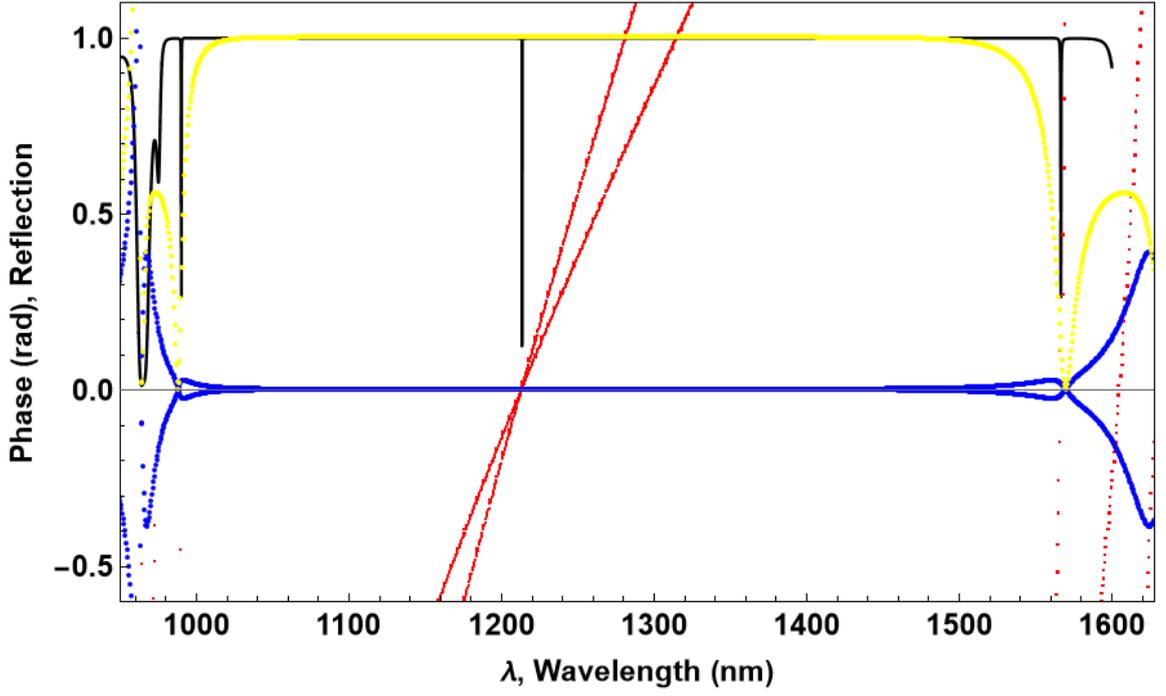

**Figure 6.** Spectra of $\text{Re}[\varphi_m - \varphi_m']$ (red), $\text{Im}[\varphi_m]$ (blue), $\xi$ (yellow) and reflection spectrum $|R|^2$ (black) for structure $(AB|D_1|AB|D_2|BA)$. The parameters of the structure are $N_1 = N_3 = 7$, $N_2 = 14$, $n_1 = 1.3$, $n_2 = 2.6$, $d_1 = 233.333$ nm, $d_2 = 116.667$ nm, $d_{d1} = 679.104$ nm, $d_{d2} = 452.736$ nm, $n_{d1} = n_{d2} = 1.34$. The normal incidence of light on the PC was considered.

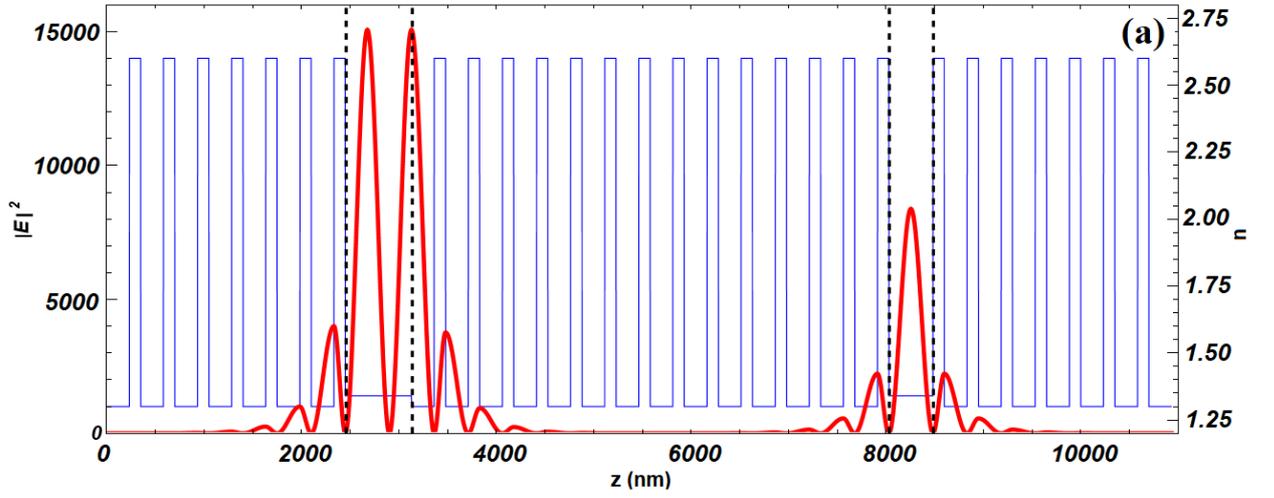

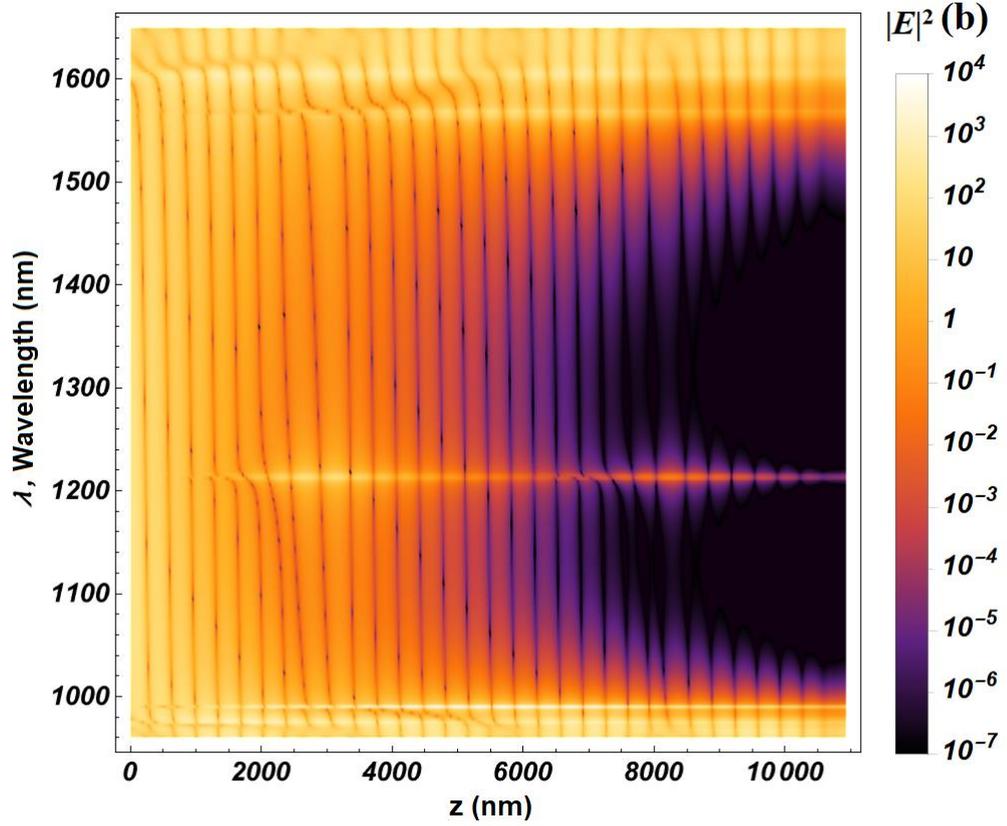

**Figure 7**. Electromagnetic field intensity $|E|^2$ inside the 1D PC with the structure (AB|D$_1$|AB|D$_2$|BA), calculated at the merging wavelength of two DMs $\lambda = 1213.33$ nm, where *z* is the coordinate inside the PC (red), and the refractive index (blue) (a); evolution spectrum of the field intensity $|E|^2$ when the wavelength $\lambda$ and the coordinate inside the structure *z* change (b). The parameters of the structure are the same as in Figure 6. Dashed lines show the boundaries between PC and DL.

Thus, by changing the parameters of the PC structure ($n_{d1}$, $n_{d2}$, $d_{d1}$, $d_{d2}$, $N_1$, $N_2$, $N_3$, etc.), it is possible to obtain a variety of structures, where the DMs will merge at a given wavelength.

**4. Derivation of analytical equations for the zero imaginary phase**

In the previous part of the paper, we have already shown the influence of the imaginary part of the total phase $\text{Im}[\varphi_m]$ from Eq. (21) on the DM amplitude in the reflection spectrum of a 1D PC with two DLs. When the imaginary phase is zero, the mode amplitude is maximized and such DM is easier to use in various applications of PC, such as optical sensors, filters and resonators. That is why we proceed to a more detailed investigation of the conditions of equality to zero of the imaginary part of the total phase to obtain the maximum DM amplitude.

Let's make the following substitutions in Eq. (21):

$$b = \frac{\tilde{r}_{\text{II}}^*}{r_{\text{III}}} e^{i\Delta\varphi} + \frac{\tilde{r}_{\text{I}}^*}{r_{\text{II}}} e^{-i\Delta\varphi} |r_{\text{II}}|^2, \quad D = \left(\frac{\tilde{r}_{\text{II}}^*}{r_{\text{III}}} e^{i\Delta\varphi} + \frac{\tilde{r}_{\text{I}}^*}{r_{\text{II}}} e^{-i\Delta\varphi} |r_{\text{II}}|^2\right)^2 - \frac{4\tilde{r}_{\text{I}}^*\tilde{r}_{\text{II}}^*}{r_{\text{II}} r_{\text{III}}}. \tag{29}$$

Taking into account the substitution (29), the Eq. (21) will take the following form:

$$\varphi_{m_{1,2}} = -i \cdot \ln\left[\frac{1}{2}\left(b \pm \sqrt{D}\right)\right] + \pi q_1. \tag{30}$$

Taking into account the expansion of the complex logarithm into real and imaginary parts, we obtain:

$$\varphi_{m_{1,2}} = -i \cdot \ln\left|\frac{1}{2}(b \pm \sqrt{D})\right| + \text{Arg}(b \pm \sqrt{D}) + \pi q_1, \tag{31}$$

For the imaginary part of $\varphi_{m_{1,2}}$ to be zero, this condition must be satisfied:

$$\left|\frac{1}{2}(b \pm \sqrt{D})\right| = 1. \tag{32}$$

Eq. (33) is equivalent to: $\sqrt{\left(\frac{1}{2}(b \pm \sqrt{D})\right)\left(\frac{1}{2}(b \pm \sqrt{D})\right)^*} = 1$, from which follows:

$$|b|^2 + |D| - 4 = \pm(b^*\sqrt{D} + b(\sqrt{D})^*), \tag{33}$$

Keeping in mind that $z + z^* = 2Re[z] = 2|z|\cos[\text{Arg}(z)]$, we get the equation:

$$\frac{|b|^2 + |D| - 4}{2|b||\sqrt{D}|} = \pm \cos\left(\text{Arg}[b] - \frac{1}{2}\text{Arg}[D]\right). \tag{34}$$

Using Eq. (34), it is possible to select such parameters of the PC structure at which the imaginary total phase equals zero at the necessary wavelength.

## 5. Numerical results for different structures

In this part of the paper, several types of 1D PC structures with two DLs are compared and the DM behavior for each of them is analyzed, and interesting numerical results are presented.

In our previous work [24], we have demonstrated the influence of the symmetry of the structure of a 1D PC with a single DL on the field distribution inside it. Similar investigations were also presented in [52-55]. Here we decided to analyze the behavior of the phase of radiation passing through the DLs, as well as the distribution of the electromagnetic field inside the PC with two DLs, and also consider the features of phase and field behavior in a perfectly symmetric PC structure. For this purpose, several different 1D PC structures with two DLs were considered, in which the reflection spectrum exhibited two distinct DMs in close proximity to one another. The following types of structures were used: the structure completely without mirror symmetry AB|D₁|AB|D₂|AB (as shown in Figure 1), the structure with a mirrored central PC AB|D₁|BA|D₂|AB, and the structure with symmetric PCs, where around each DL there is a layer with a higher refractive index: AB|D₁B|AB|D₂|BA. For all structures, the first order DM was considered. Also, the same DLs were used in the PC structures. The parameters of the structure are as follows: $N_1 = N_2 = N_3 = 5$, $n_1 = 1.5$, $n_2 = 2.4$, $d_1 = 307.7$ nm, $d_2 = 192.3$ nm, $n_{d1} = n_{d2} = 1.33$. The normal incidence of light on the PC was considered. For each structure, we plotted the real and imaginary total phase spectra $\varphi_m$ the reflection spectrum $|R|^2$, as well as the electromagnetic field intensity distribution within the structure at wavelengths corresponding to the DM position, and evolution plots of the electromagnetic field intensity dependence on the radiation wavelength $\lambda$ and the coordinate inside the structure $z$. Our calculations showed that the field distributions plotted at the wavelengths of the first and second DMs are almost identical, so in all the examples below (except for the evolutionary plots) the wavelength of the shortwave DM. The corresponding figures and their analysis are given below.

1) AB|D₁|AB|D₂|AB structure:

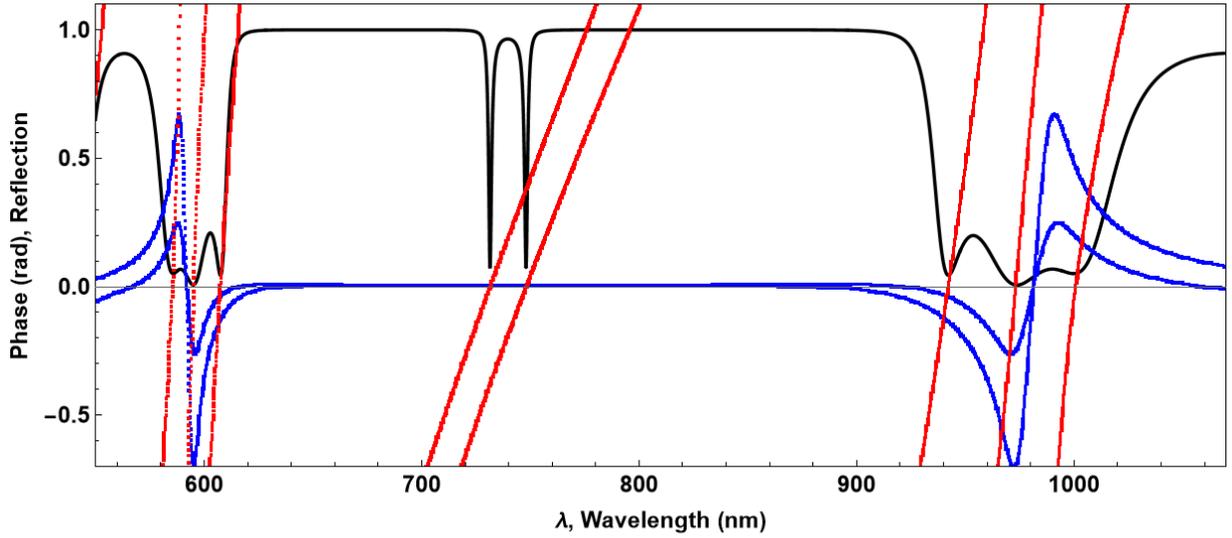

**Figure 8**. Spectra of $\text{Re}[\varphi_m - \varphi_m']$ (red), $\text{Im}[\varphi_m]$ (blue) and reflection spectrum $|R|^2$ (black) for structure (AB|D₁|AB|D₂|AB). The parameters of the structure are $N_1 = N_2 = N_3 = 5$, $n_1 = 1.5$, $n_2 = 2.4$, $d_1 = 123$ nm, $d_2 = 77$ nm, $d_{d1} = d_{d2} = 140$ nm, $n_{d1} = n_{d2} = 1.33$. The normal incidence of light on the PC was considered.

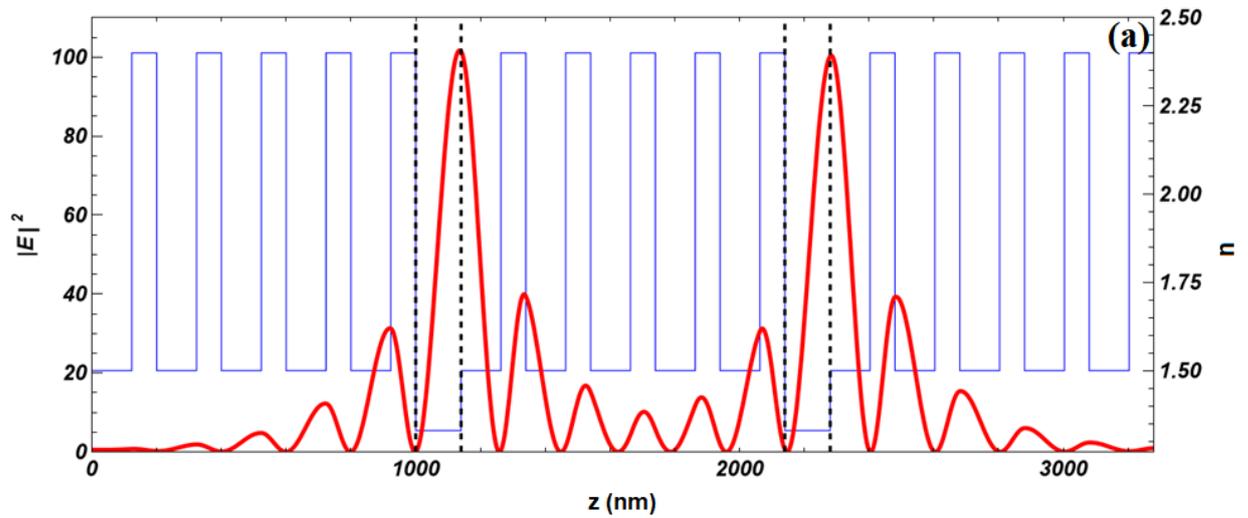

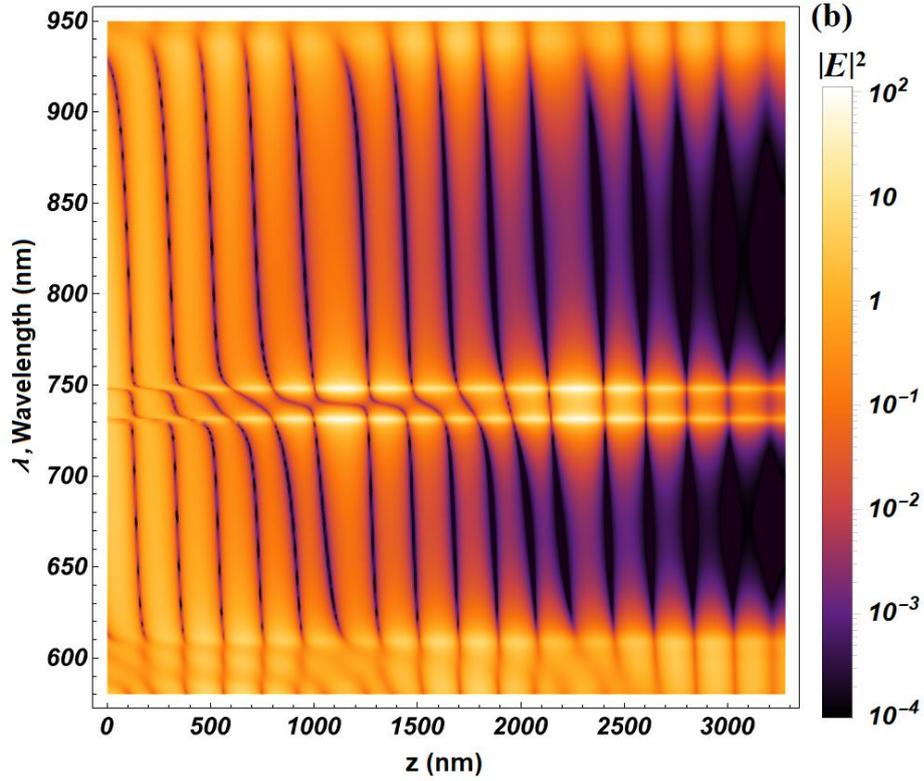

**Figure 9**. Electromagnetic field intensity $|E|^2$ inside the 1D PC with the structure (AB|D$_1$|AB|D$_2$|AB), calculated at the wavelength $\lambda_1 = 748.1$ nm, where z is the coordinate inside the PC (red), and the refractive index (blue) (a); evolution spectrum of the field intensity $|E|^2$ when the wavelength $\lambda$ and the coordinate inside the structure $z$ change (b). The parameters of the structure are the same as in Figure 8. Dashed lines show the boundaries between PC and DL.

2) AB|D$_1$|BA|D$_2$|AB structure:

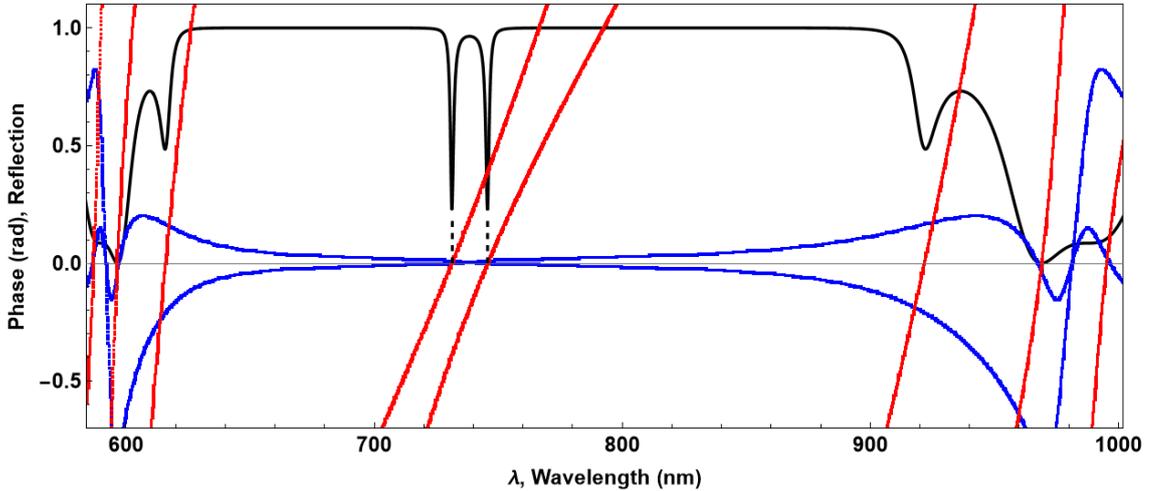

**Figure 10**. Spectra of $\text{Re}[\varphi_m - \varphi_m']$ (red), $\text{Im}[\varphi_m]$ (blue) and reflection spectrum $|R|^2$ (black) for structure (AB|D$_1$|BA|D$_2$|AB). The parameters of the structure are $N_1 = N_2 = N_3 = 5$, $n_1 = 1.5$, $n_2 = 2.4$, $d_1 = 123$ nm, $d_2 = 77$ nm, $d_{d1} = d_{d2} = 277$ nm, $n_{d1} = n_{d2} = 1.33$. The normal incidence of light on the PC was considered.

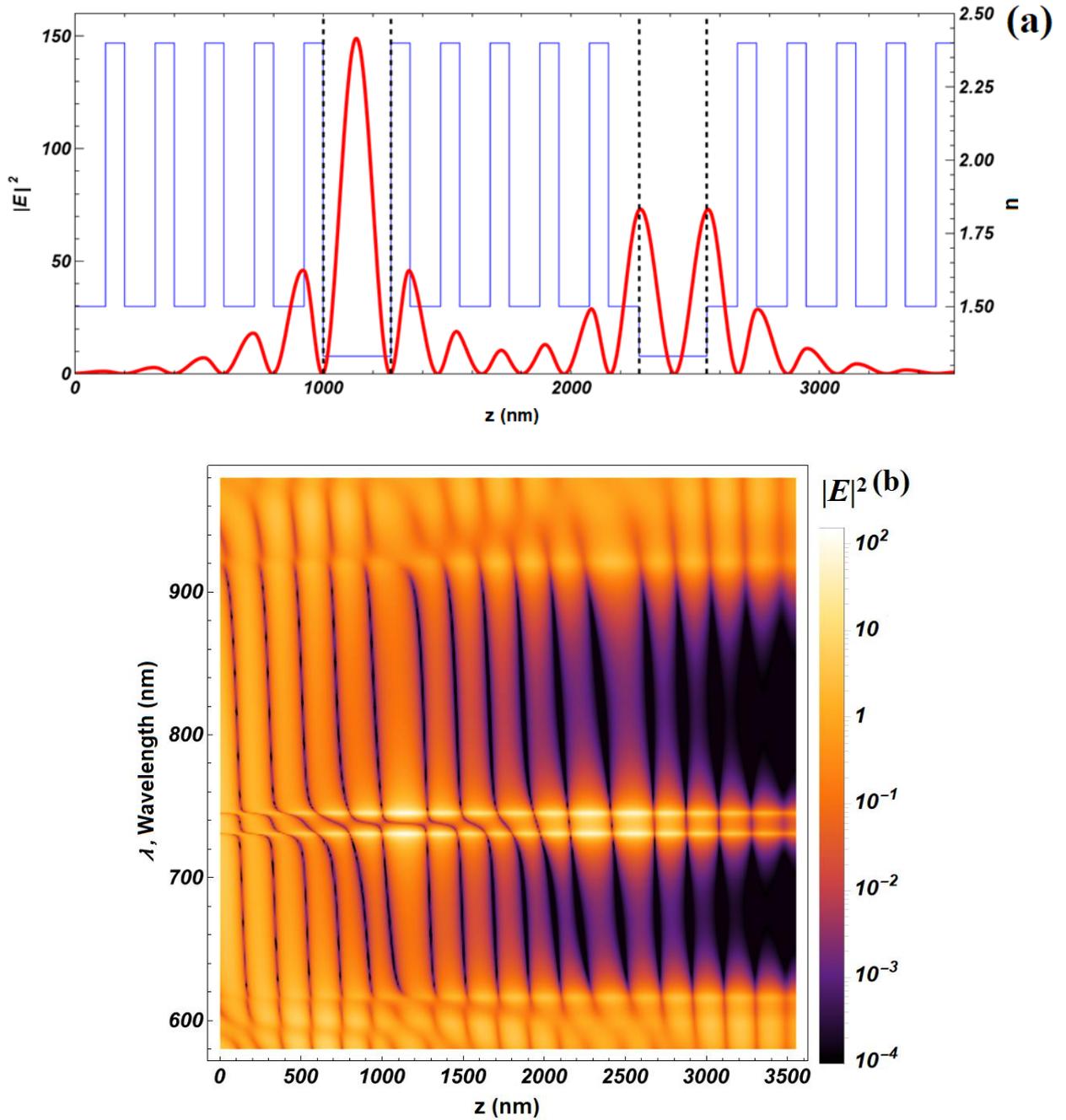

**Figure 11**. Electromagnetic field intensity $|E|^2$ inside the 1D PC with the structure (AB|D$_1$|BA|D$_2$|AB), calculated at the wavelength $\lambda_1 = 728$ nm, where z is the coordinate inside the PC (red), and the refractive index (blue) (a); evolution spectrum of the field intensity $|E|^2$ when the wavelength λ and the coordinate inside the structure z change (b). The parameters of the structure are the same as in Figure 10. Dashed lines show the boundaries between PC and DL.

3) AB|D₁B|AB|D₂|BA structure:

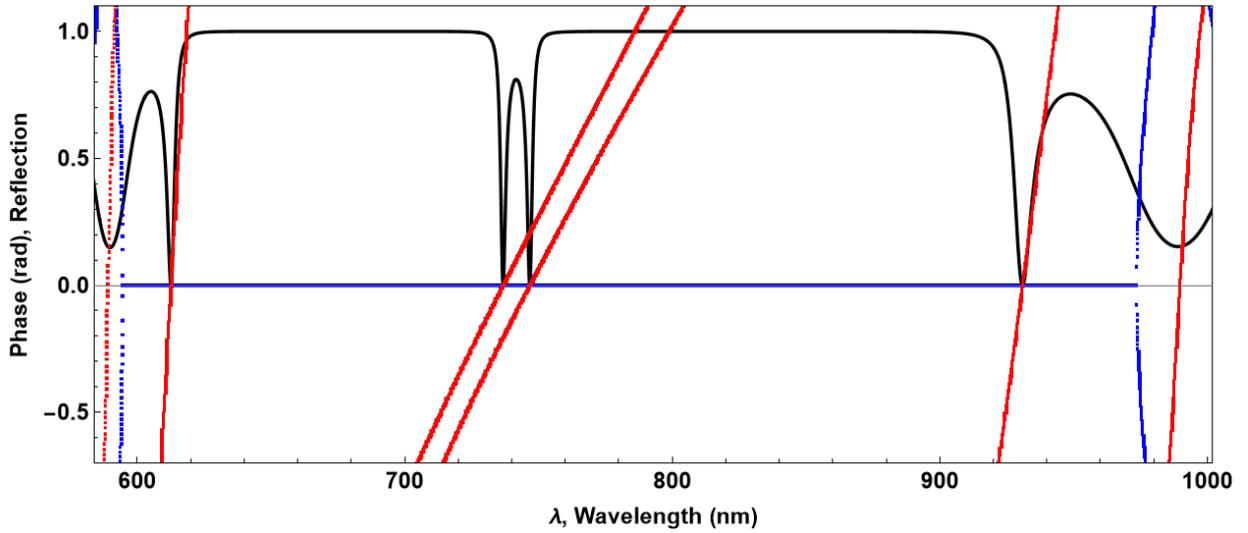

**Figure 12**. Spectra of $\mathrm{Re}[\varphi_m - \varphi_m']$ (red), $\mathrm{Im}[\varphi_m]$ (blue) and reflection spectrum $|R|^2$ (black) for structure (AB|D₁B|AB|D₂|BA). The parameters of the structure are $N_1 = N_2 = N_3 = 5$, $n_1 = 1.5$, $n_2 = 2.4$, $d_1 = 123$ nm, $d_2 = 77$ nm, $d_{d1} = d_{d2} = 281$ nm, $n_{d1} = n_{d2} = 1.33$. The normal incidence of light on the PC was considered.

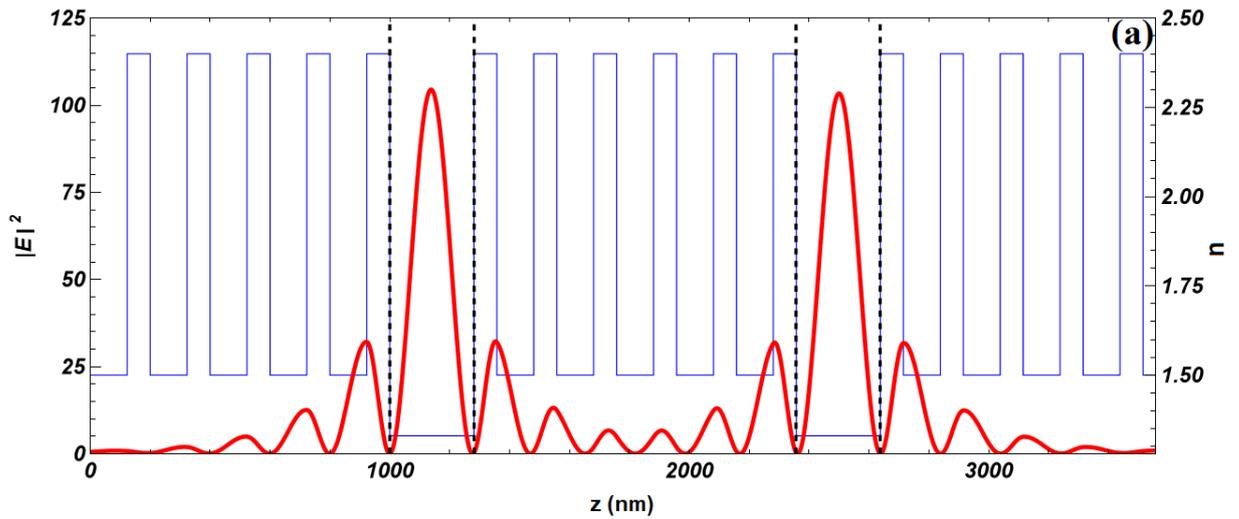

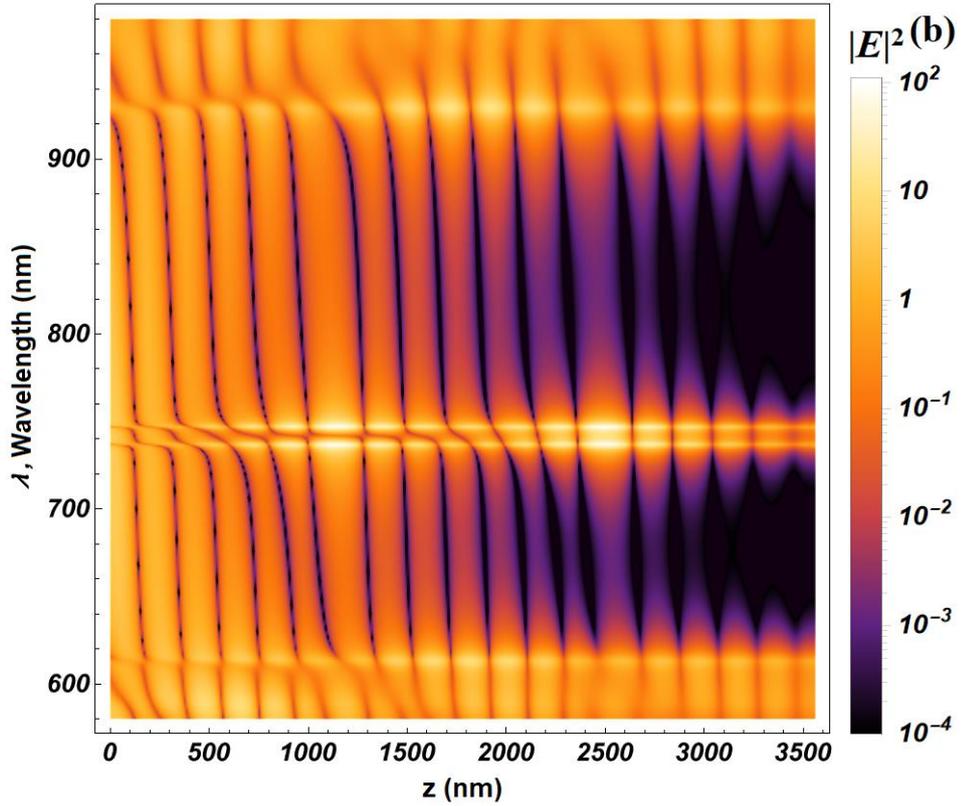

**Figure 13**. Electromagnetic field intensity $|E|^2$ inside the 1D PC with the structure (AB|D$_1$|BA|D$_2$|AB), calculated at the wavelength $\lambda_1 = 737$ nm, where z is the coordinate inside the PC (red), and the refractive index (blue) (a); evolution spectrum of the field intensity $|E|^2$ when the wavelength λ and the coordinate inside the structure z change (b). The parameters of the structure are the same as in Figure 12. Dashed lines show the boundaries between PC and DL.

A comparison of Figures 8 and 12 reveals that in these structures, the imaginary total phase Im[$\varphi_m$] assumes values close to zero throughout the entire region of the first PBG. However, in the first case, both solutions behave symmetrically and take negative values outside the PBG. In addition, the value of the imaginary total phase is not exactly zero, due to which the amplitude of the DMs is not maximized. And in the second case, a sharp jump of the imaginary phase occurs at the edges of the PBG, and then the solutions behave asymmetrically. In this case, the imaginary part of the total phase is exactly equal to zero, due to which the amplitude of the DMs is maximal in the whole region inside the first PBG. In Figure 10, the imaginary total phase Im[$\varphi_m$] assumes values close to zero only near the PBG center. In contrast, its value is moving away from zero as one approaches the edges of the PBG. For all considered structures, the real phase Re[$\varphi_m$] is zero at all wavelengths values corresponding to the minima of the reflection spectrum $|R|^2$. Figures 9a, 11a, and 13a demonstrate that electromagnetic field localization occurs in both DLs within the PC.

In Figures 9a and 13a, the field distributions near the two DLs are the same because the DLs have the similar thickness and boundary layers. But in the first case, the minimum of electric field intensity is localized at the first edge of each DL, while the maximum is localized at the second edge. And for the second case, the field intensity has a maximum in the center of each DLs, while at the edges of the DLs it has a minimum. In Figure 11a, the field maximum regions for both DLs are not identical to each other: the maximum is localized at the center of the first DL, while the minimum is localized at the center of the second DL. For this type of structure, at each DM order, the field maximum will be observed at the center of one of the DLs, while the field minimum will be observed at the center of the other DL. At the edges of the first DL field has a minimum, while at the edges of the second DL it has a maximum. In this case, our calculations show that for each considered structure with each increase in the order of the DMs there is an increase in the number of local field maxima inside each of the DLs by 1. Also, from Figures 9b, 11b and 13b it can be seen that the field takes the maximum value at two wavelengths corresponding to each of the DMs within the first PBG. From Figures 9b and 11b it can be seen, for the first wavelength, the number of field maxima between the two DLs is 5, while for the second wavelength it is 4. And in Figure 13b, for the first wavelength, the number of field localization maxima between the two DLs is 6, while for the second wavelength it is 5.

In conclusion of this section, it can be stated that the different symmetry of PCs with two DLs with other identical parameters has a significant impact on the behavior of the phase of optical radiation passing through the DLs, particularly its imaginary component. Additionally, it influences the distribution of the electromagnetic field within and around DLs. The symmetric distribution of the field within two DLs is observed in the case of the PC without mirror symmetry AB|$D_1$|AB|$D_2$|AB and in the perfectly symmetric PC AB|$D_1$B|AB|$D_2$|BA, while in the structure with a mirrored center AB|$D_1$|BA|$D_2$|AB there will be no symmetric field distribution: in the center of one of the DLs the field maximum will be observed, and in the center of the other - the minimum. For the structure with perfectly symmetric PCs, the imaginary part of the total phase within the entire first PBG is identically equal to zero. Due to this, the amplitude of the DMs is maximized, which makes such a structure the most convenient choice for a variety of technological applications, including filters, sensors and others.

## 6. Conclusion

In conclusion, analytical expressions for the position of DMs appearing within the PBG in 1D PCs with two DLs were obtained for the first time from the condition of zero reflection from the entire structure. In addition, it was shown that the problem of finding DMs of a PC with two DLs can be reduced to two independent problems for two PCs, each with a single defect, in the case of weak coupling between DLs, that means when reflection coefficient of the middle mirror $r_{II}$ is close to unity. Also, the condition of merging two DMs in the case of non-zero coupling between DLs was derived. Additionally, a condition on the DM with a zero value of the reflection coefficient was also obtained. Several typical 1D PC structures with two DLs have been compared, and the DM behavior and the electromagnetic field intensity distribution within the first PBG of such structures have been analyzed for each of them. It is shown that different symmetry of PCs with two DLs with other identical parameters strongly affects the behavior of the phase of optical radiation passing through DLs (especially its imaginary part), as well as the distribution of the electromagnetic field within and around DLs. The distribution of the electromagnetic field intensity within the first PBG for several cases of DM merging for different orders of DMs was investigated, and the peculiarities of the PC structures, in which very large and also very small values of the field intensity localized on the DLs of the PC are achieved, were shown. The results of this work allow to considerably simplify the analysis and optimization of optical sensors and filters based on 1D PCs with two DLs and allow to better understand numerous numerical results obtained earlier by other researchers.

**Acknowledgment**

The work was supported by the Russian Science Foundation (Grant № 24-22-00283).

**Data availability statement**

The data that support the findings of this study are available upon request from the authors.

**Conflicts of Interest**

No potential conflict of interest was reported by the authors.